\documentclass[journal=jacsat,manuscript=article]{achemso}
\setkeys{acs}{maxauthors=99}
\setkeys{acs}{etalmode=truncate}

\usepackage[utf8]{inputenc}
\usepackage{graphicx}
\usepackage{booktabs}
\usepackage{caption} 
\usepackage{threeparttable}  
\usepackage{hyperref}
\usepackage{todonotes}
\usepackage{siunitx}
\usepackage{lineno}
\usepackage{lipsum}
\usepackage{tabularx}
\usepackage{multirow}

\newcolumntype{C}{>{\centering\arraybackslash}X}
\newcolumntype{a}{>{\hsize=.2\hsize}C}
\newcolumntype{b}{>{\hsize=.2\hsize}C}
\newcolumntype{c}{>{\hsize=.4\hsize}C}
\newcolumntype{d}{>{\hsize=.2\hsize}C}

\usepackage[version=3]{mhchem} 
\usepackage{verbatim}

\author{Song Yin}
\altaffiliation{These authors contributed to the work equally.}
\affiliation[University of Illinois at Urbana-Champaign]
{Department of Chemical and Biomolecular Engineering, University of Illinois Urbana-Champaign, Urbana, IL 61801, United States}

\author{Xuenan Mi}
\altaffiliation{These authors contributed to the work equally.}
\affiliation[University of Illinois at Urbana-Champaign]
{Center for Biophysics and Quantitative Biology, University of Illinois Urbana-Champaign, Urbana, IL 61801, United States}

\author{Diwakar Shukla}
\affiliation[University of Illinois Urbana-Champaign]
{Department of Chemical and Biomolecular Engineering, University of Illinois Urbana-Champaign, Urbana, IL 61801, United States}
\alsoaffiliation{Center for Biophysics and Quantitative Biology, University of Illinois Urbana-Champaign, Urbana, IL 61801, United States}
\alsoaffiliation{Department of Bioengineering, University of Illinois Urbana-Champaign, Urbana, IL 61801, United States}
\email{diwakar@illinois.edu}

\title[An \textsf{achemso} demo]
  {Leveraging Machine Learning Models for Peptide-Protein Interaction Prediction}

\abbreviations{IR,NMR,UV}

\DeclareUnicodeCharacter{2212}{-}
\begin{document}

\begin{abstract}
Peptides play a pivotal role in a wide range of biological activities through participating in up to 40\% protein-protein interactions in cellular processes. They also demonstrate remarkable specificity and efficacy, making them promising candidates for drug development. However, predicting peptide-protein complexes by traditional computational approaches, such as Docking and Molecular Dynamics simulations, still remains a challenge due to high computational cost, flexible nature of peptides, and limited structural information of peptide-protein complexes. In recent years, the surge of available biological data has given rise to the development of an increasing number of machine learning models for predicting peptide-protein interactions. These models offer efficient solutions to address the challenges associated with traditional computational approaches. Furthermore, they offer enhanced accuracy, robustness, and interpretability in their predictive outcomes. This review presents a comprehensive overview of machine learning and deep learning models that have emerged in recent years for the prediction of peptide-protein interactions.
\end{abstract}


\section{Introduction}
Peptides consist of short chains of amino acids connected by peptide bonds, typically comprising 2 to 50 amino acids. One of the most critical functions of peptides is their mediation of 15-40\% of protein-protein interactions (PPIs) \cite{London2013}. PPIs play essential roles in various biological processes within living organisms, including DNA replication, DNA transcription, catalyzing metabolic reactions and regulating cellular signal \cite{Peng2016}. Peptides have become promising drug candidates due to their ability to modulate PPIs. Over the past century, Food and Drug Administration (FDA) has approved more than 80 peptide drugs \cite{Muttenthaler2021}, with insulin being the pioneering therapeutic peptide used extensively in diabetes treatment. Compared with the small molecules, peptide drugs demonstrate high specificity and efficacy \cite{Wang2022}. Additionally, compared with other classes of drug candidates, peptides have more flexible backbones, enabling their better membrane permeability \cite{Wang2022}. 

Rational design of peptide drugs is challenging and costly, due to the lack of stability and the big pool of potential target candidates. Therefore, computational methodologies that have proven effective in small molecule drug design have been adapted for modelling peptide-protein interactions (PepPIs). These computational techniques include Docking, Molecular Dynamics (MD) simulations, and machine learning (ML) and deep learning (DL) models. Docking approaches enable exploration of peptide binding positions and poses in atomistic details, facilitating the prediction of binding affinities \cite{Meng2011, Wang2019,Charitou2022,Lensink2016,Lensink2020}. However, peptides are inherently flexible and they can interact with proteins in various conformations. These conformations often change during the binding process \cite{Ciemny2018}. MD simulation is another approach to model the peptide-protein interaction. The peptide-protein binding and unbinding process can be studied thermodynamically and kinetically through MD simulations \cite{Paul2017,Morrone2017,Morrone2017Jan,Kilburg2018,Wang2019ChemicalReview,Zou2020,Zalewski2021,Chen2022,Ciemny2018}. But sampling the complex energy landscapes associated with peptide-protein interactions typically requires intensive computational resources and time. The accuracy of Docking and MD simulations both rely on the knowledge of protein structures, therefore the limited availability of peptide-protein complex structures has restricted the utility of these two approaches. 

In recent years, ML and DL models have been widely used in the field of computer-aided drug design. These models offer an alternative way to address the inherent challenges associated with Docking and MD simulations in modeling PepPIs. Due to the large amount of available biological data, many ML/DL models are routinely employed to obtain sequence-function relationship, achieving comparable predictive performance to structure-based models. This is because sequence data contains evolutionary, structural and functional information across protein space. Furthermore, compared with Docking and MD simulation, ML/DL models exhibit greater efficiency and generalizability. Trained ML/DL models are capable of predicting PepPIs in a single pass, but it's hard to do large-scale docking and MD simulations due to their resource-intensive and time-consuming nature. Moreover, with the development of interpretable models, DL models are no longer regarded as black boxes; they can provide valuable insights into residue-level contributions to peptide-protein binding predictions. 

\begin{figure}[!htbp]
    \centering
    \includegraphics[width=\textwidth]{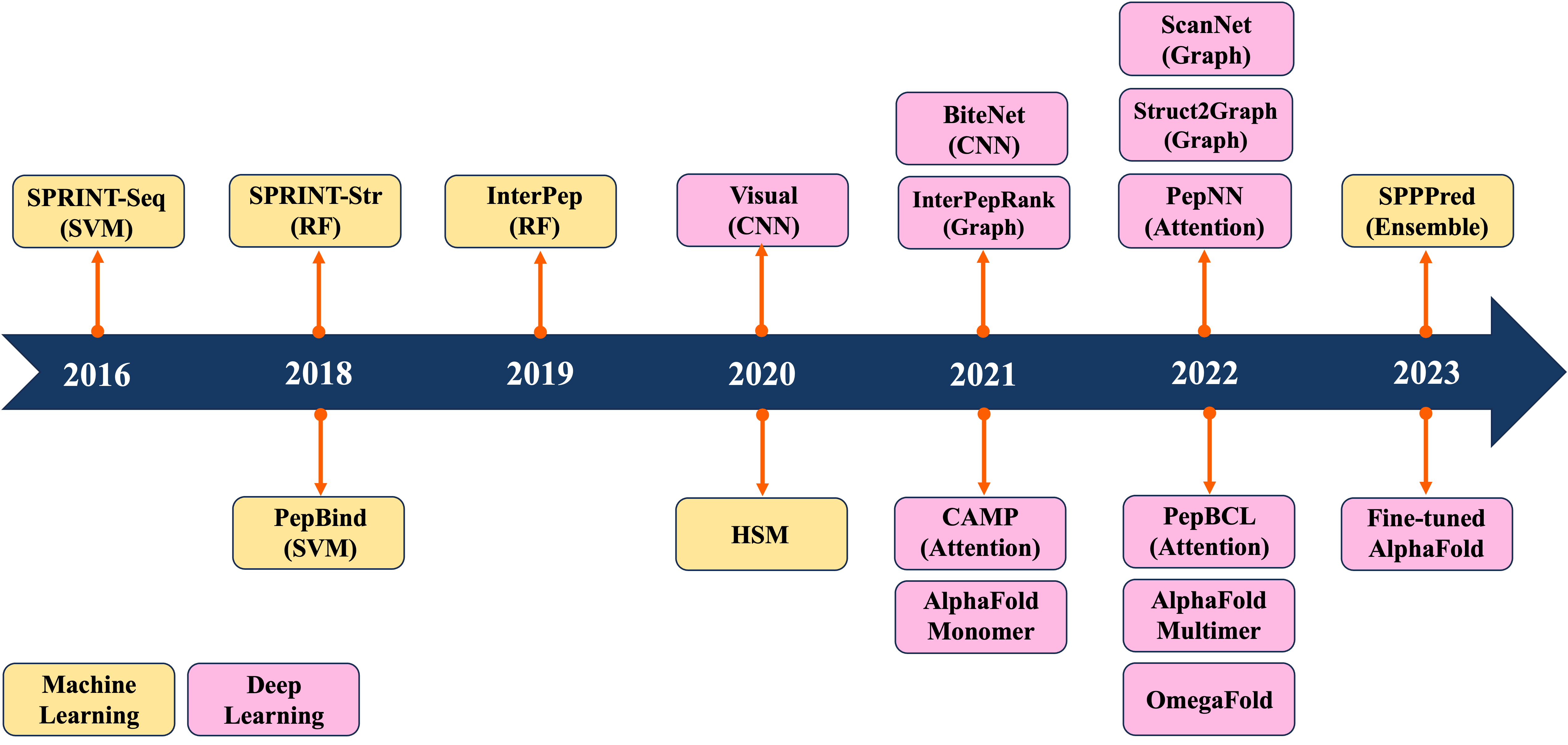}
    \caption{Timeline of recent Machine Learning and Deep Learning methods for PepPIs prediction.}
    \label{fig:1}
\end{figure}

Previous reviews mainly summarized ML/DL models for predicting PPIs \cite{Zhang2017, Casadio2022, Soleymani2022, Hu2022, Lee2023,Tang2023}. They have traditionally categorized computational methods for predicting PPIs into two main classes: sequence-based and structure-based approaches. Sequence-based methods extract information only from sequence data, whereas structure-based methods rely on the information derived from peptide-protein complex structures. Recently, ML/DL models have increasingly integrated both sequence and structure information to enhance their predictive performance. In this review, we systematically summarize the progress made in predicting PepPIs. From ML perspective, we include Support Vector Machine (SVM) and Random Forest (RF). ML models typically require manual feature extraction from sequence and structure datasets. But DL models, including Convolutional Neural Network (CNN), Graph Convolutional Network (GCN) and Transformer, automatically extract multi-layer feature representations from data. To the best of our knowledge, this is the first review to summarize the ML/DL work for specifically predicting PepPIs. Figure \ref{fig:1} shows the timeline illustrating the evolution of ML/DL methods in the context of PepPIs predictions. Table \ref{table:1} summarizes the details of ML/DL models discussed in this review.

\begin{table}[!htbp]
    \caption{Overview of Machine Learning models for PepPIs prediction}
    \label{table:1}
    \scriptsize
    \begin{tabularx}{\textwidth}{aabcd}
 	\hline
 	\textbf{Model name} & \textbf{Baseline Model} & \textbf{Data Type $\&$ Datasets} & \textbf{Key Ideas} & \textbf{Model performance} \\ 
 	\hline
	SPRINT-Seq\cite{Taherzadeh2016} & SVM & Protein sequences from BioLip\cite{Yang2012} Protein Sequence  & First ML model predicted PepPIs only based on sequence features & ACC$\colon 0.66$, AUC$\colon 0.71$, MCC$\colon 0.33$, SEN$\colon 0.64$, SP$\colon0.68$\\ 
	\hline
	PepBind\cite{Zhao2018} & SVM & Protein sequences from BioLip\cite{Yang2012}  & Intrinsic disorder-based features were first introduced & AUC$\colon 0.76$, MCC$\colon 0.33$, SEN$\colon 0.32$, PRE$\colon0.45$\\
	\hline
	SPRINT-Str\cite{Taherzadeh2017} & RF & Protein–peptide complex sequences $\&$ structures from BioLip\cite{Yang2012} &  Used structural information and employed the RF classifier & ACC$\colon 0.94$, AUC$\colon 0.78$, MCC$\colon 0.29$, SEN$\colon 0.24$, SP$\colon0.98$ \\
	\hline
	InterPep\cite{Johanssonkhe2019} & RF & Protein–peptide complex structures from RCSB PDB\cite{Berman2000} & Predicted what region of the protein structure the peptide is most likely to bind &  ACC$\colon 0.81$, SEN$\colon 0.51$\\
	\hline
	SPPPred\cite{Shafiee2023} & Ensemble$\colon$ SVM, RF, KNN & Protein sequences from BioLip database\cite{Yang2012} & Ensemble learning model was applied for effectively handling imbalanced dataset & ACC$\colon 0.95$, AUC$\colon 0.71$, MCC$\colon 0.23$, F1$\colon 0.31$,SEN$\colon 0.32$, SP$\colon0.96$\\ 
	\hline
	Hierarchical Statistical Mechanical (HSM) \cite{Cunningham2020} & HSM & Peptide binding domain (PBD)–peptide structures from UniProt\cite{uniprot2018} & Introduced bespoke HSM model to predict the affinities of peptide binding domain (PBD)–peptide interactions & AUC$\colon 0.97$ (PBD$\colon$ PDZ) \\
	\hline
	Visual\cite{Wardah2020} & CNN & Protein sequences from BioLip\cite{Yang2012} & Protein sequence features were transformed into images and CNN was first applied to predict PepPIs & AUC$\colon 0.73$, MCC$\colon 0.17$, SEN$\colon 0.67$, SP$\colon0.69$\\
	\hline
	BiteNet${_P}_p$\cite{Kozlovskii2021} & CNN & Protein–peptide complex structures from BioLip\cite{Yang2012} & Utilized 3D CNN and protein structures directly to predict protein–peptide binding sites & AUC$\colon 0.91$, MCC$\colon 0.49$, PRE$\colon 0.53$ \\
	\hline
	InterPepRank\cite{Johanssonkhe2021} & GCN & Protein–peptide complex structures from RCSB PDB\cite{Berman2000}  &  Achieve high accuracy in predicting both binding sites and conformations for disordered peptides & AUC$\colon 0.86$   \\
	\hline
	ScanNet\cite{Tubiana2021} & Geometric DL Architecture & Protein–peptide complex structures from Dockground\cite{Kundrotas2017} & An end-to-end, interpretable geometric DL model that learns features directly from 3D structures & ACC$\colon 0.88$, AUC$\colon 0.69$, SEN$\colon 0.50$, PRE$\colon 0.74$\\
	\hline
	Struct2Graph\cite{Baranwal2022} & GCN and Attention & Protein–peptide complex structures from IntAct\cite{Orchard2013}, STRING\cite{Szklarczyk2018}, and UniProt\cite{uniprot2018} &  A GCN-based mutual attention classifier accurately predicting interactions between query proteins exclusively from 3D structural data & ACC$\colon 0.99$, AUC$\colon 0.99$, MCC$\colon 0.98$, F1$\colon 0.99$, SEN$\colon 0.98$, SP$\colon 0.99$, PRE$\colon 0.99$, NPV$\colon 0.98$\\  
	\hline
	CAMP\cite{Lei2021} & CNN and self-attention & Protein-peptide complexes sequences from RCSB PDB\cite{Berman2000} $\&$ DrugBank\cite{Wishart2017} & Took account of sequence information of both protein and peptide, and identified binding residues of peptides & AUC$\colon 0.87$, AUPR$\colon 0.64$\\
	\hline
	PepNN\cite{Abdin2022} & Transformer & Protein-peptide complexes sequences and structures from RCSB PDB\cite{Berman2000} &  Utilized a multi-head reciprocal attention layer to update the embeddings of both peptide and protein; Transfer learning was applied to solve the limited protein-peptide complex structures issue & AUC$\colon 0.86$, MCC$\colon 0.41$\\
	\hline
	 \end{tabularx}
	\end{table}

\begin{table}[!htbp]
    \ContinuedFloat
    \caption{Overview of Machine Learning models for PepPIs prediction (continued)}
    \label{table:1}
    \scriptsize
    \begin{tabularx}{\textwidth}{aabcd}
 	\hline
 	\textbf{Model name} & \textbf{Baseline Model} & \textbf{Data Type $\&$ Datasets} & \textbf{Key Ideas} & \textbf{Model performance} \\ 
 	\hline
	PepBCL\cite{Wang2022PEPBCL} & BERT-based contrastive learning framework & Protein sequences from BioLip database\cite{Yang2012} & An end-to-end predictive model; Contrastive learning module was used to tackle imbalanced data issue & AUC$\colon 0.82$, MCC$\colon 0.39$, SEN$\colon 0.32$, SP$\colon0.98$, PRE$\colon0.54$\\
	\hline
	AlphaFold Monomer\cite{Jumper2021, Tsaban2022, Shanker2023} &  MSA based transformer  & Protein sequences $\&$ structures from Uniclust30\cite{Mirdita2016} and RCSB PDB\cite{Berman2000} & \multirow{2}{=}{\centering{Adding the peptide sequence \textit{via} a poly-glycine linker to the C-terminus of the receptor monomer sequence could mimic peptide docking as monomer folding}}  & SR$\colon 0.75$ (within 1.5 $\AA$ RMSD) in Tsaban\textit{et al.} \cite{Tsaban2022} $\&$ SR$\colon 0.33$ (Fraction of Native Contacts = $0.8$ as cutoff) in Shanker\textit{et al.}\cite{Shanker2023}  \\ \cline{1-3} \cline{5-5}
	OmegaFold\cite{Wu2022, Shanker2023} & Protein language model & Protein sequences $\&$ structures from Uniref50\cite{Suzek2014}, RSCB PDB\cite{Berman2000}, CASP\cite{Weissenow2022}, and CAMEO\cite{Robin2021} &  & SR$\colon 0.20$ (Fraction of Native Contacts = $0.8$ as cutoff) in Shanker\textit{et al.}\cite{Shanker2023} \\ \cline{1-3} \cline{5-5}
	\hline
	AlphaFold Multimer\cite{Evans2021, Shanker2023} &  MSA based transformer &  Protein complexes sequences $\&$ structures from RSCB PDB\cite{Berman2000} and Benchmark 2\cite{Ghani2021} & Improved the accuracy of predicted multimeric interfaces between two or more proteins & SR$\colon 0.53$ (Fraction of Native Contacts = 0.8 as cutoff) in Shanker\textit{et al.}\cite{Shanker2023}  \\
	\hline
	Fine-tuned AlphaFold\cite{Motmaen2023} & MSA based transformer & Peptide-MHC complex structures RSCB PDB\cite{Berman2000} & Leveraging and fine-tuning AF2 with existing peptide-protein binding data could improve its PepPIs predictions &  AUC$\colon 0.97$ (Class \MakeUppercase{\romannumeral 1}) $\&$ AUC$\colon 0.93$ (Class \MakeUppercase{\romannumeral 2})  \\
	\hline
    \end{tabularx}
     \begin{tablenotes}
      \small
      \item \textit{Abbreviations}: ACC: Accuracy;  AUC: Area under ROC curve; AUPR: Area under precision-recall curve; MCC: Matthews correlation coefficient; SEN: Sensitivity; SP: Specificity; PRE: Precision; SR: Success Rate.
    \end{tablenotes}
\end{table}


\section{Machine Learning Models for Peptide-Protein Interactions Prediction}
\textbf{Support Vector Machine (SVM)}. SVM is a powerful ML algorithm commonly employed for classification tasks. The objective of SVM is to determine the optimal hyperplane that effectively separates data points belonging to different classes in the feature space. The selection criteria for this optimal hyperplane aims to maximize the margins between the closest points of distinct classes, thereby minimizing misclassification rates.  

\begin{figure}[!htbp]
    \centering
   \includegraphics[width=\textwidth]{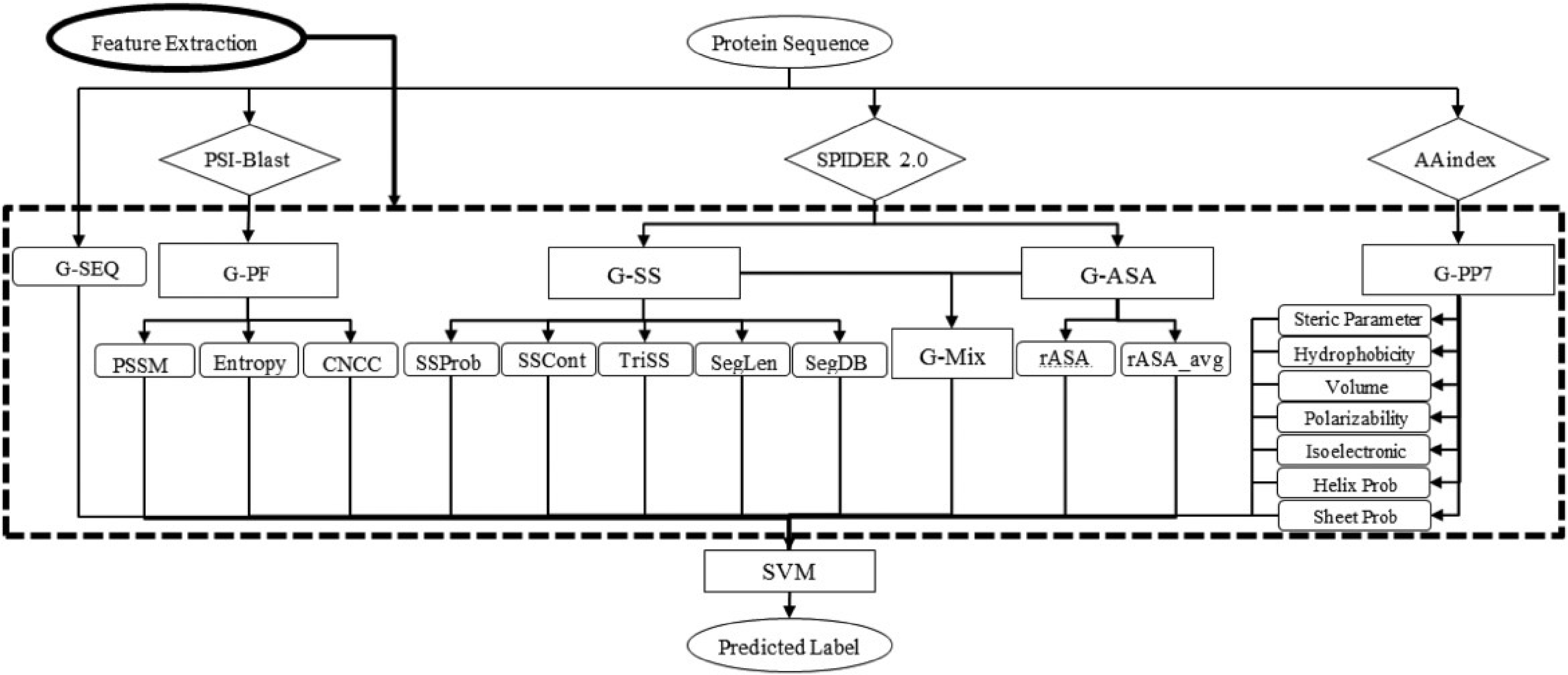}
   \caption{The input features and architecture of SPRINT-Seq. G-SEQ: sequence feature. G-PF: Sequence profile from Position Specific Scoring Matrix (PSSM). G-SS: Secondary Structure-based features. G-ASA:Accessible Surface Area-based features. G-PP7: Physicochemical-based feature group. Adapted with permission from G.Taherzadeh, Y.Yang, T.Zhang, A.W.-C.Liew and Y.Zhou, \textit{Journal of Computational Chemistry}, 2016, \textbf{37}, 1223–1229. Copyright 2024 John Wiley and Sons.}
   \label{fig:2}       
\end{figure}

SPRINT-Seq (Sequence-based prediction of Protein–peptide Residue-level INTeraction sites) is the first ML based prediction of peptide-protein binding sites only using sequence features \cite{Taherzadeh2016}. Various types of information were extracted from protein sequence to create a feature dataset, including one-hot encoded protein sequences, evolutionary information \cite{Altschul1997}, predicted accessible surface area \cite{Heffernan2015}, secondary structure \cite{Heffernan2015}, and physiochemical properties \cite{Meiler2001}.These features were fed into a classification model, SVM, to predict the label for each residue (Figure \ref{fig:2}). SPRINT-Seq yielded Matthews’ Correlation Coefficient (MCC) of 0.326, sensitivity of 0.64 and specificity of 0.68 on an independent test set. The importance of each feature was also evaluated, the most crucial feature distinguishing binding from non-binding residues is the sequence evolution profile. This sequence-based technique's performance is comparable or better than structure-based models (Peptimap \cite{Lavi2013}, Pepite \cite{Petsalaki2009}, PinUp \cite{Liang2006}, VisGrid \cite{Li2008}) for peptide-binding sites prediction.

To improve the accuracy of sequence-based prediction, Zhao \textit{et al.} introduced intrinsic disorder as a feature within sequence representation \cite{Zhao2018}. Peptides that participate in peptide-protein interactions exhibit consistent attributes of short linear motifs, primarily found in the intrinsic disordered regions (IDRs). These attributes include short length, flexible structure and weak binding affinity \cite{Weatheritt2012}.  In addition to the novel sequence representation, they designed a consensus-based method called PepBind \cite{Zhao2018}. This method combines SVM classification model with the template-based methods S-SITE and TM-SITE \cite{Yang2013}. The aggregation of these three individual predictors yielded better performance than all three individual methods and outperformed the first sequence-based method SPRINT-Seq. 

\textbf{Random Forest (RF)}. RF is another supervised ML algorithm for classification and regression, which combines multiple decision trees to create a ``forest". During training of a RF for classification, each tree contributes a vote. The forest subsequently selects the classification with the majority of votes as the predicted outcome. All decision trees comprising the RF are independent models. While individual decision trees may contain errors, the collective majority vote of the ensemble ensures more robust and accurate predictions, thereby enhancing the reliability of RF predicted results.

A RF model, SPRINT-Str \cite{Taherzadeh2017} (Structure-based Prediction of Residue-level INTeraction), was developed to predict the putative peptide-protein binding residues and binding sites by combining both sequence-based and structure-based information. The sequence information in the input includes Position Specific Scoring Matrix (PSSM) for all amino acids in the protein and entropy calculated based on PSSM. Structural information includes Accessible Surface Area (ASA) calculated by DSSP (Define Secondary Structure of Proteins)\cite{Kabsch1983}, Secondary Structure (SS) calculated by DSSP,\cite{Kabsch1983} half-sphere exposure (HSE) representing the solvent exposure using residue contact numbers in upward and downward hemispheres along with pseudo C$\beta$–C$\alpha$ bond,\cite{Hamelryck2005} and flexibility calculated by iModeS\cite{LpezBlanco2014} to describe the functional motions of proteins.\cite{Dykeman2010} A RF classifier was further trained and tested to predict the binding residues. The Density-based Spatial Clustering of Applications with Noise (DBSCAN) algorithm \cite{osti_421283} was then applied to cluster spatially neighboring binding site residues. The largest cluster was selected as the predicted binding site with a corresponding reliability score. SPRINT-Str achieved robust performance in predicting binding residues with MCC of 0.293 as well as Area Under the Receiver Operating Characteristic Curve (ROC AUC) of 0.782. For instance, when testing the model's performance on peptide binding with the human tyrosine phosphatase protein PTPN4 PDZ domain (PDBID: 3NFK) \cite{Babault2011}, 15 out of 17 binding residues were correctly predicted, and the predicted binding sites were similar to the actual binding sites. SPRINT-Str is one of the representative ML models that pass structural features into the models and achieves remarkable success in predicting PepPIs. 

The structures of proteins or peptide-protein complexes can also be directly used as input to ML models. The underlying premise of this approach is that, if a PepPI shares similarities with a certain interaction surface, that well-characterized surface can serve as a template for modeling other PepPIs. The InterPep model \cite{Johanssonkhe2019} constructs four steps to better represent this idea: Mass Structural Alignment (MSA), Feature Extraction, RF Classification, and Clustering. A Template Modeling (TM) score larger than 0.5 was used to screen out candidate templates. Overall, InterPep accurately predicted 255 out of 502 (50.7\%) binding sites for the top 1 prediction and correctly identified 348 out of 502 (69.3\%) binding sites within the top 5 predictions, which demonstrates it's a useful tool for the identification of peptide-binding sites.

\textbf{Ensemble Learning}. In the pursuit of a more robust predictive model for protein-peptide binding sites, Shafiee \textit{et al.} adopted an ensemble-based ML classifier named SPPPred \cite{Shafiee2023}. Ensemble learning stands out as an effective strategy for handling imbalanced datasets, as it allows multiple models to collectively contribute to predictions, resulting in enhanced robustness, reduced variance, and improved generalization \cite{CamachoGmez2021}. 

In the SPPPred algorithm, the ensemble learning technique of bagging \cite{Polikar2006} was employed to predict peptide binding residues. The initial step in bagging involves generating various subsets of data through random sampling with replacement, a process known as bootstrapping. For each bootstrap dataset, distinct classification models are trained, including Support Vector Machine (SVM), K-Nearest Neighbors (KNN), and Random Forest (RF). Subsequently, for each residue, the class with the majority of votes across these models is determined as the final predicted label. This ensemble method consistently demonstrates strong and comparable performance on independent test sets, with F1 score of 0.31, accuracy of 0.95, MCC of 0.23. 

\textbf{Other State-Of-The-Art (SOTA) Models}. There are some SOTA bespoke ML models that achieve great success for the predictions of PepPIs, for example, Hierarchical statistical mechanical modeling (HSM).\cite{Cunningham2020} A dataset of 8 peptide-binding domain (PBD) families was applied to train and test the HSM model, including PDZ, SH2, SH3, WW, WH1, PTB, TK, and PTP, which cover 39\% of human PBDs. The HSM model defines a pseudo-Hamiltonian, which is a machine-learned approximation of Hamiltonian that maps the system state to its energy\cite{AlQuraishi2014}. The predicted PepPI probability is derived from the sum of pseudo-Hamiltonian corresponding to each PBD-peptide sequence pair. In total, 9 models were developed, including 8 separate HSM/ID models (ID means independent domain, one for each protein family) and a single unified HSM/D model covering all families (D means domains). The HSM model remarkably outperformed other ML models such as NetPhorest\cite{Miller2008} and PepInt\cite{Kundu2014}. By computing the energies from pseudo-Hamiltonian, the HSM model can evaluate and rank the possibilities of different PepPI patterns, facilitating the verification of existing PepPI ensembles and the discovery of new possible PepPI ensembles. Furthermore, the HSM model provides detailed explanations of the peptide-protein binding mechanism, demonstrating a strong interpretability. Using peptide binding with HCK-SH3 domain (PDBID: 2OI3) \cite{Schmidt2007} as an example, the HSM model gave a detailed examination and explanation of the peptide-SH3 domain binding mechanism. The “W114 tryptophan switch” binding motif \cite{FernandezBallester2004} was correctly recognized by the HSM model. Additionally, a conserved triplet of aromatic residues W114-Y132-Y87 was previously identified as contributing to the peptide binding with the HCK-SH3 domain\cite{Lee1995, Zarrinpar2003}. However, the HSM model also found that Y89 and Y127 had similar predicted energetic profiles as W114, suggesting a new possible W-Y-Y aromatic triplet. By mapping the predicted interaction energies to the complex structure, the HSM model successfully recognized the repulsive binding regions and attractive binding regions. The predicted attractive binding interface correctly aligns with the previously studied RT-loop and proline recognition pocket\cite{Lee1995, Zarrinpar2003}, demonstrating the strong predictive and interpretative ability of the HSM model.


\section{Deep Learning Models for Peptide-Protein Interactions Prediction}

\textbf{Convolutional Neural Network (CNN)}. CNN is a class of neural networks that have demonstrated the great success in processing image data \cite{https://doi.org/10.48550/arxiv.1511.08458}. The design of CNN was inspired by biological visual system in humans. When humans see an image, each neuron in the brain processes information within its own receptive filed and connects with other neurons in a way to cover the entire image. Similarly, each neuron in a CNN also only processes data in its receptive field. This approach allows CNNs to dissect simpler patterns initially and subsequently assemble them into more complex patterns. A typical CNN architecture consists of three layers: the convolutional layer, the pooling layer, and the fully connected layer. In the convolutional layer, a dot product is computed between two matrices—the first being a kernel with a set of learnable parameters, and the second representing a portion of the receptive field. The kernel slides across the entire image, generating a two-dimensional representation. The pooling layer replaces the output of the convolutional layer at each location by deriving a summary statistic of the nearby outputs. This serves to reduce the size of the feature maps, subsequently decreasing training time. Finally, the fully connected layer connects the information extracted from the previous layers to the output layer and eventually classify the input into a label. The biological data could be transformed into an image-like pattern, therefore CNN could be applied to binding site identification. 

\begin{figure}[!htbp]
   \centering
   \includegraphics[width=\textwidth]{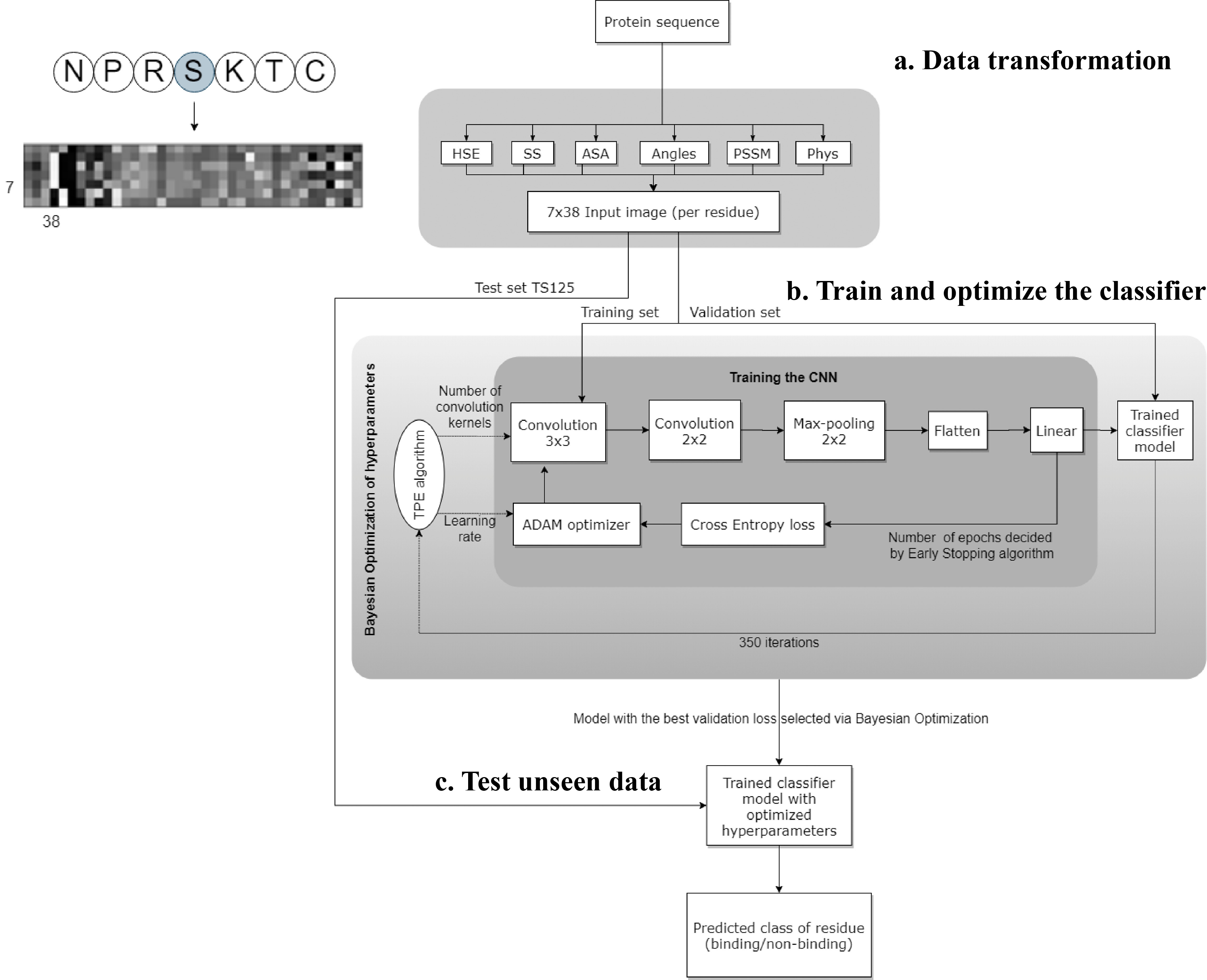}
   \caption{The workflow of Visual model. (a) Transforming protein sequence into 7$\times$38 input image (per residue). In order from left to right of image: 3 pixels represents Half Sphere Exposure (HSE) \cite{Hamelryck2005}, 3 pixels represent the predicted probabilities of different secondary structure, 1 pixel represents the Accessible Surface Area (ASA) value, 4 pixels represent the local backbone angles, 20 pixels represent the Position Specific Scoring Matrix (PSSM), and 7 pixels represent the physicochemical properties of the amino acids. (b) Training and optimizing hyperparamters of CNN. (c) Testing the optimized CNN on unseen test data to predict the label of each residue (binding/non-binding). Adapted with permission from W. Wardah, A. Dehzangi, G. Taherzadeh, M. A. Rashid, M. Khan, T. Tsunoda and A. Sharma, \textit{Journal of Theoretical Biology}, 2020, \textbf{496}, 110278. Copyright 2024 Elsevier.}
   \label{fig:4}       
\end{figure}

Wardah \textit{et al.} applied CNNs for identifying peptide-binding sites by introducing a CNN-based method named, Visual \cite{Wardah2020}. In Visual algorithm, features were extracted from protein sequence, like HSE \cite{Hamelryck2005}, secondary structure \cite{Yang2016}, ASA \cite{Yang2016}, local backbone angles \cite{Yang2016}, PSSM \cite{Altschul1997} and  physicochemical properties \cite{Kawashima1999}.  These features were stacked horizontally resulting in a feature vector with a length of 38. Visual employs a sliding window approach to capture the local context of each residue. For a given residue, the feature vectors of the three upstream and three downstream residues were combined into a matrix, resulting in a 2-dimensional array with size of 7$\times$38. An illustrative example of the input data in an image-like format is depicted in Figure \ref{fig:4}, showcasing the center residue Serine (S) within a window size of 7. A 7$\times$38 image is generated as input of CNN classifier. The Visual model comprises two sets of convolutional layers, followed by a pooling layer and a fully connected layer (Figure \ref{fig:4}).  Visual was applied to identify the peptide binding sites of protein and achieved sensitivity of 0.67 and ROC AUC of 0.73. 

BiteNet${_P}_p$ \cite{Kozlovskii2021} is another CNN-based model that converts 3D protein structures to 4D tensor-based representations and feeds them into a 3D CNN to learn the probability of PepPIs and predict the peptide binding sites/domain. The 4D tensor has the first three dimensions corresponding to the x, y, and z dimensions, and the fourth dimension corresponding to 11 channels including atomic densities of 11 different atom types such as aromatic carbon, sulfur, amide nitrogen, carbonyl oxygen, and so forth. These four-dimensional tensor-based representations were then fed into 10 three-dimensional convolutional layers to obtain the probability score of “hot spots”, which are determined as the geometric centers of each segmented peptide-protein interface. This model outperforms SOTA methods with ROC AUC of 0.91 and MCC of 0.49. The model showed promising power for the prediction of peptide-protein binding sites, but the model's performance is limited by the input protein orientation and sensitivity to the protein conformations. Therefore, BiteNet${_P}_p$ could be improved by using representations that could handle the protein rotation invariance. 

\textbf{Graph Convolutional Network (GCN)}. Graph based models have been widely used to illustrate the PPIs and PepPIs based on the peptide/protein structures \cite{Johanssonkhe2021, Baranwal2022, Tubiana2021, fout2017protein, Cao2020, Gao2023, Huang2023, Rau2022}. Graph embedding \cite{SanchezLengeling2021} includes nodes (vertices) representing different entities and edges (links) representing the relationships between them. For proteins, graphs typically assign amino acids and related information as nodes, with the distances and connections between amino acids represented as edges. This approach allows for the direct observation of information from protein 3D structures without involving hand-crafted features.\cite{Wieder2020, Tang2023} GCNs \cite{Zhang2019, Zhou2020} are a type of neural network that can be used to learn graph embeddings. Similar to CNNs, GCNs take graph embeddings as input and progressively transform them through a series of localized convolutional and pooling layers where each layer updates all vertex features. The updated embeddings are passed through a classification layer to obtain the final classification results.\cite{SanchezLengeling2021, Zhang2019} GCNs have been successfully applied to protein binding site prediction, with models such as PipGCN \cite{fout2017protein} and EGCN \cite{Cao2020} achieving great success. More recently, a number of GCN-based models have also been applied for PepPIs prediction. 

InterPepRank \cite{Johanssonkhe2021} is a representative GCN that has been developed to predict the PepPIs. In this model, billions of decoys (computational protein folding structure) were generated by the PIPER \cite{Kozakov2006} docking tool as the training and testing set, respectively. The peptide-protein complexes were then represented as graphs with one-hot encoded nodes illustrating individual residues, PSSM \cite{Remmert2011}, self-entropy,\cite{Remmert2011} and one-hot encoded edges denoting the residue interactions. Both node and edge features were then passed through edge convolution layers with the output from each layer concatenated and fed into a global pooling layer and two dense layers to predict the LRMSD (ligand root-mean-square deviation) of decoys. InterPepRank achieved a median ROC AUC of 0.86, outperforming other benchmarking methods such as PIPER,\cite{Kozakov2006} pyDock3,\cite{Cheng2007} and Zrank.\cite{Pierce2007} For example, in the case of a fragment from the center of troponin I (peptide) binding with the C-terminal domain of Akazara scallop troponin C (receptor),\cite{Agirrezabala2011} the peptide was proved to be disordered when unbound and become an ordered $\alpha$-helical structure upon binding,\cite{Basu2017} following the induced-fit binding mechanism. Predicting the peptide binding conformation and binding sites for systems with induced-fit mechanisms is extremely challenging.  The top 100 decoys predicted by both InterPepRank and Zrank showed that both methods can find the true binding site of the peptide. However, InterPepRank achieved an accuracy of 96\% in predicting the peptide as an $\alpha$-helical structure, while Zrank only achieved an accuracy of less than 50\%, where half of the peptide decoys' secondary structures were predicted as either random coils or $\beta$-sheets. Therefore, InterPepRank is a powerful tool for predicting both binding sites and conformations, even in cases where the peptide is disordered when unbound. This is a significant advantage over other benchmarked energy-based docking methods, which may struggle with disordered structures that are more energetically favorable in unbound states or easier to fit into false positive binding sites. 

\begin{figure}[!htbp]
   \centering
   \includegraphics[width=\textwidth]{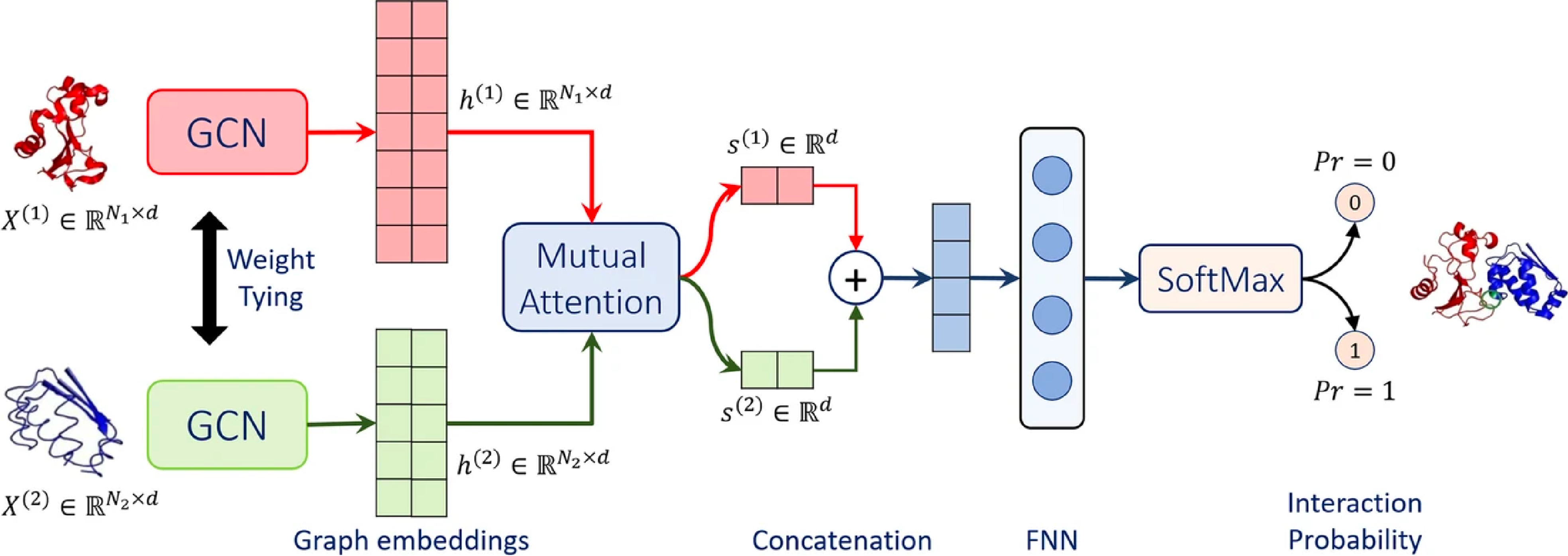}
   \caption{Struct2Graph model architecture. Struct2Graph model loads graph embeddings of both components into two weight sharing graph convolutional networks (GCNs) seperately. GCNs outputs are integrated into a mutual attention network to predict the probability of PPI and the interaction sites. Adapted with permission from M. Baranwal, A. Magner, J. Saldinger, E. S. Turali-Emre, P. Elvati, S. Kozarekar, J. S. VanEpps, N. A. Kotov, A. Violi and A. O. Hero, \textit{BMC Bioinformatics}, 2022, \textbf{23}, 370. This article is licensed under a Creative Commons Attribution 4.0 International License, permitting unrestricted reproduction and adaptation provided proper crediting to author and source. Copyright 2024 Springer Nature.}
   \label{fig:5}       
\end{figure}

Struct2Graph \cite{Baranwal2022} is a novel multi-layer mutual graph attention convolutional network for structure-based predictions of PPIs  (Figure \ref{fig:5}). Coarse-grained graph embeddings were generated by two GCNs with weight sharing for both components of the protein complexes. These embeddings were then passed through a mutual attention network to extract the relevant features for both proteins and concatenated into a single embedding vector. By calculating attention weights, residues with large learned attention weights are more important and more likely to contribute towards interaction. The vector was further passed into a feed-forward network (FFN) and a final Softmax layer to get the probability for PPI. Struct2Graph outperformed the feature-based ML models and other SOTA sequence-based DL models, achieving an accuracy of 98.89\% on positive/negative samples balanced dataset, and accuracy of 99.42\% on a positive/negative samples unbalanced dataset (positive:negative = 1:10). Residue-level interpretation was conducted to identify the residues' contribution to PepPIs. For example, \textit{Staphylococcus aureus} Phenol Soluble Modulins (PSMs) peptide PSM$\alpha _{1}$ \cite{TayebFligelman2017} competes with high mobility group box-1 protein (HMGB1) to bind with toll-like receptor-4 (TLR4),\cite{Wang2017} thus inhibiting HMGB1-mediated phosphorylation of NF-$\kappa$B.\cite{Chu2018} For the PSM$\alpha _{1}$-TLR4 complex, Struct2Graph demonstrated impressive accuracy of 92\%, and the predicted binding residues aligned with the previously identified TLR4 active binding sites.  Notably, peptide residues 2Gly and 10Val were accurately predicted as the peptide binding residues. Furthermore, Struct2Graph’s predictions corroborated the previously studied competitive binding mechanism,  indicating that both PSM$\alpha _{1}$ peptide and HMGB1 bind to the same area of TLR4. 

\begin{figure}[!htbp]
   \centering
   \includegraphics[scale = 0.60]{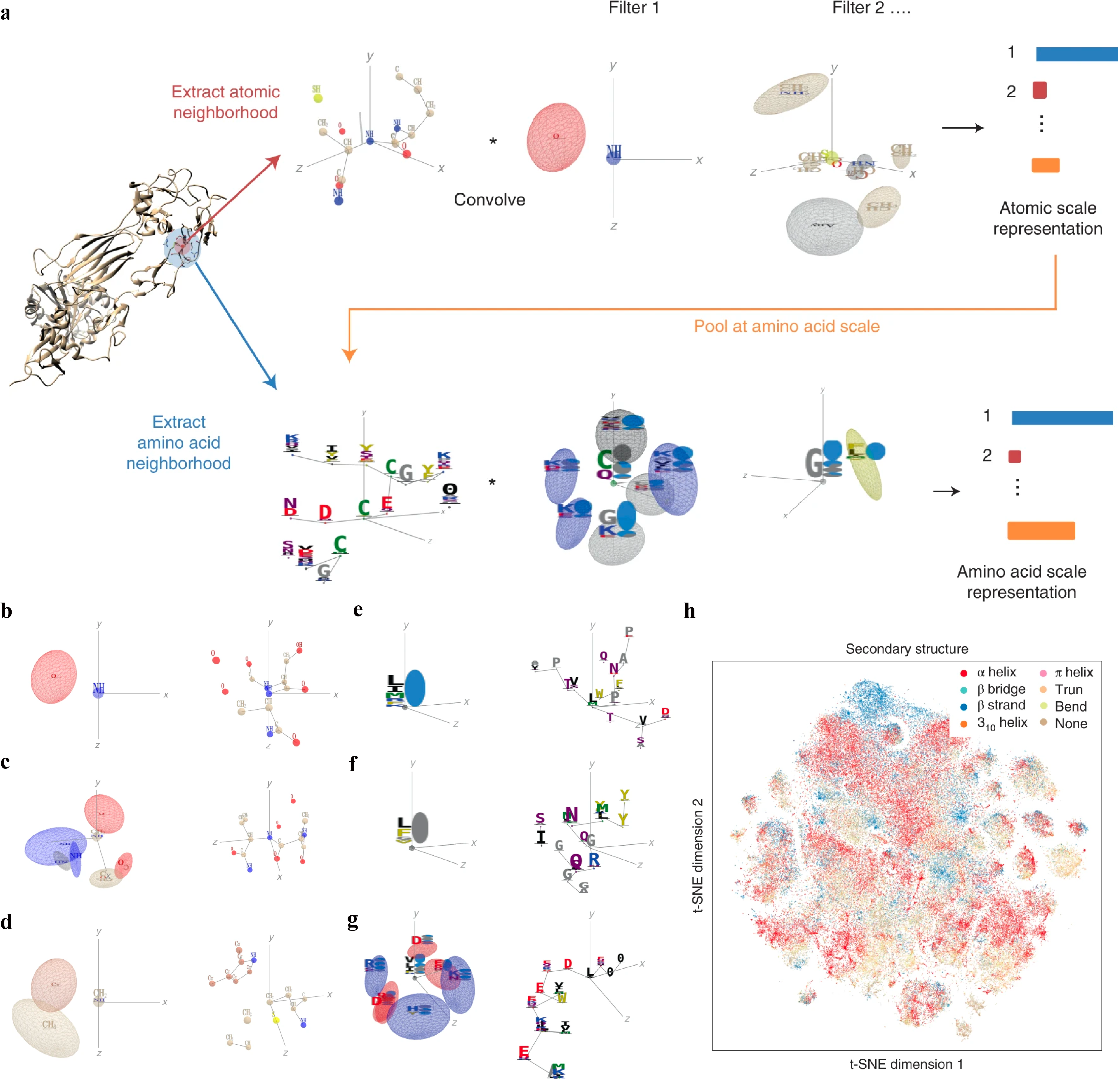}
   \caption{(a) Overview of the ScanNet model architecture. Point cloud including neighboring atoms information was first extracted for each atom from the protein structure. Point cloud was then passed through linear filters to detect specific atom interaction patterns, and yielding an atomic-scale representation. This representation was pooled to amino acid scale, concatenated with the extracted neighboring amino acid attributes from the protein structure, and then applied to similar procedure as before to identify amino acid neighborhood and representations. (b-f) Each panel shows one learned atom-level spatio-chemical patterns on the left and corresponding top-activating neighborhood on the right. (b) N-H-O hydrogen bond, (c) two oxygen atoms and three NH groups in a specific arrangement, (d) a carbon in the vicinity of a methyl group and an aromatic ring. (e-g) Each panel shows one learned amino acid-level spatio-chemical pattern on the left and one corresponding top-activating neighborhood on the right. (e) solvent-exposed residues, positively correlated with the output probability (r=0.31), (f) buried hydrophobic amino acids, negatively correlated with the output probability (r=−0.32), (g) The hotspot O-ring architecture, exposed hydrophobic amino acid surrounded by exposed, charged amino acids, positively correlated with the output probability (r=0.29). (h) Two-dimensional projection on secondary structure of the learned amino acid scale representation using t-SNE. Reproduced with permission from J. Tubiana, D. Schneidman-Duhovny and H. J. Wolfson, \textit{bioRxiv}, 2021. This article is licensed under a CC BY 4.0 International License, permitting unrestricted reproduction and adaptation provided proper crediting to author and source. Copyright 2024 Cold Spring Harbor Laboratory.}
   \label{fig:6}       
\end{figure}

Interpretable DL graph models have also been employed for the PepPI predictions. Recently, an end-to-end geometric DL architecture known as ScanNet (Spatio-chemical arrangement of neighbors neural NETwork),\cite{Tubiana2021} was developed that integrated multi-scale spatio-chemical arrangement information of atoms, amino acid, along with multiple sequence alignment (MSA) for detecting protein–protein binding sites (PPBS). The model took the protein sequence, tertiary structure, and optionally position-weight matrix from MSA of evolutionarily related proteins as input. It first extracted all the atomic neighborhood embeddings, which were then passed through several filters to learn the atomic scale representations. To further reduce the dimensions, atom-wise representations were pooled at the amino acid scale, mixed with extracted amino acid information, and fed into trainable filters to yield amino acid scale representations (Figure \ref{fig:6}a). With these representations containing multi-scale spatio-chemical information, ScanNet was trained for the prediction of PPBS on 20k proteins with annotated binding sites. When compared with the traditional ML method XGBoost with handcrafted features, and designed pipeline based on structural homology, ScanNet achieved the highest accuracy of 87.7\%. While the structural homology baseline performed almost the same as ScanNet, the accuracy dropped quickly when meeting with the unseen fold during the test because of its strong dependence on the homology that was previously developed. Therefore, it's crucial to understand what ScanNet has actually learned. Specifically, does the network only memorize the training data, or does it really understand the underlying protein-protein binding principles? Detailed visualization and interpretation were explored to illustrate the learned atom-wise representations and amino acid-wise representations. The network has learned different atomic patterns, such as N-H-O hydrogen bond (Figure \ref{fig:6}b), SH or NH$_2$ side-chain hydrogen donor surrounded by oxygen atoms (Figure \ref{fig:6}c), a carbon in the vicinity of a methyl group and an aromatic ring (Figure \ref{fig:6}d), and so on. The detected pattern with solvent-exposed residues frequently appearing in the protein-protein interface (Figure \ref{fig:6}e), such as Arginine (R), was positively correlated with the output probability of PPBS. However, that with the buried hydrophobic amino acids (Figure \ref{fig:6}f), such as Phenylalanine (F), was negatively correlated with the output probability of PPBS. Interestingly, the pattern with exposed hydrophobic amino acid surrounded by charged amino acids, which is the hotspot O-ring \cite{Bogan1998} architecture in protein interfaces, was positively correlated with the output probability (Figure \ref{fig:6}g). 2D t-distributed stochastic neighbor embedding (t-SNE) projections further verified that the model has already learned various amino acid-level structural features. 2D t-SNE projections on secondary structures (Figure \ref{fig:6}h) clearly illustrated that the model has learned the secondary structural information of the training complexes. With the multi-level knowledge of protein structures, ScanNet captures the underlying chemical principles of protein-protein binding. This SOTA interpretable DL model aids in a deeper understanding of PepPIs and PPIs. 

\textbf{Attention based models}. Recurrent neural networks (RNN) and long short-term memory (LSTM) are most common models for language modeling and machine translation \cite{https://doi.org/10.48550/arxiv.1808.03314}. But both RNN and LSTM suffer from the issue of handling long range dependencies, in other words they become ineffective when there is a significant gap between relevant information and the point where it is needed.  The attention mechanism was introduced to address this limitation, which enables the modeling of dependencies without being constrained by their distance in input or output sequences \cite{https://doi.org/10.48550/arxiv.1706.03762}. Attention mechanism is one of the most important developments in natural language processing. Vaswani \textit{et al.} introduced a new form of attention, called self-attention, which relates different positions of a single sequence to obtain a representation of the sequence \cite{https://doi.org/10.48550/arxiv.1808.03314}. A new architectural class, Transformer, was conceived, primarily based on the self-attention mechanism \cite{https://doi.org/10.48550/arxiv.1706.03762}. Transformer consists of multiple encoders and decoders with self-attention layers. The self-attention layer allows transformer model to process all input words at once and model the relationship between all words in a sentence. Transformer architecture led to the development of a new language model, called Bidirectional Encoder Representations from Transformers (BERT) \cite{https://doi.org/10.48550/arxiv.1810.04805}. BERT is designed to pre-train deep bidirectional representations from unlabeled text. It utilizes a ``masked language model" (MLM) objective, where some tokens from the input are randomly masked, and the model is trained to predict the masked word based on its context from both directions. Numerous deep learning architectures have emerged, either directly employing self-attention mechanisms or drawing inspiration from the Transformer architecture. These advancements have also been applied forward in predicting PepPIs.

Existing ML and DL models for predicting peptide-protein binding sites mainly focus on identifying binding residues on the protein surface. Sequence-based methods typically take protein sequences as inputs, assuming that a protein maintains fixed binding residues across different peptide binders. However, this assumption doesn't hold true for most cellular processes, as various peptides may interact with distinct protein residues to carry out diverse functions. Structure-based methods would require a target protein structure and a peptide sequence, thus limiting their applicability to proteins with available structural data. A novel DL framework for peptide-protein binding prediction was proposed, called CAMP \cite{Lei2021}, to address the above limitations. CAMP takes account of information from sequence of both peptides and target proteins, and also detect crucial binding residues of peptides for peptide drug discovery.

\begin{figure}[!htbp]
   \centering
   \includegraphics[width=\textwidth]{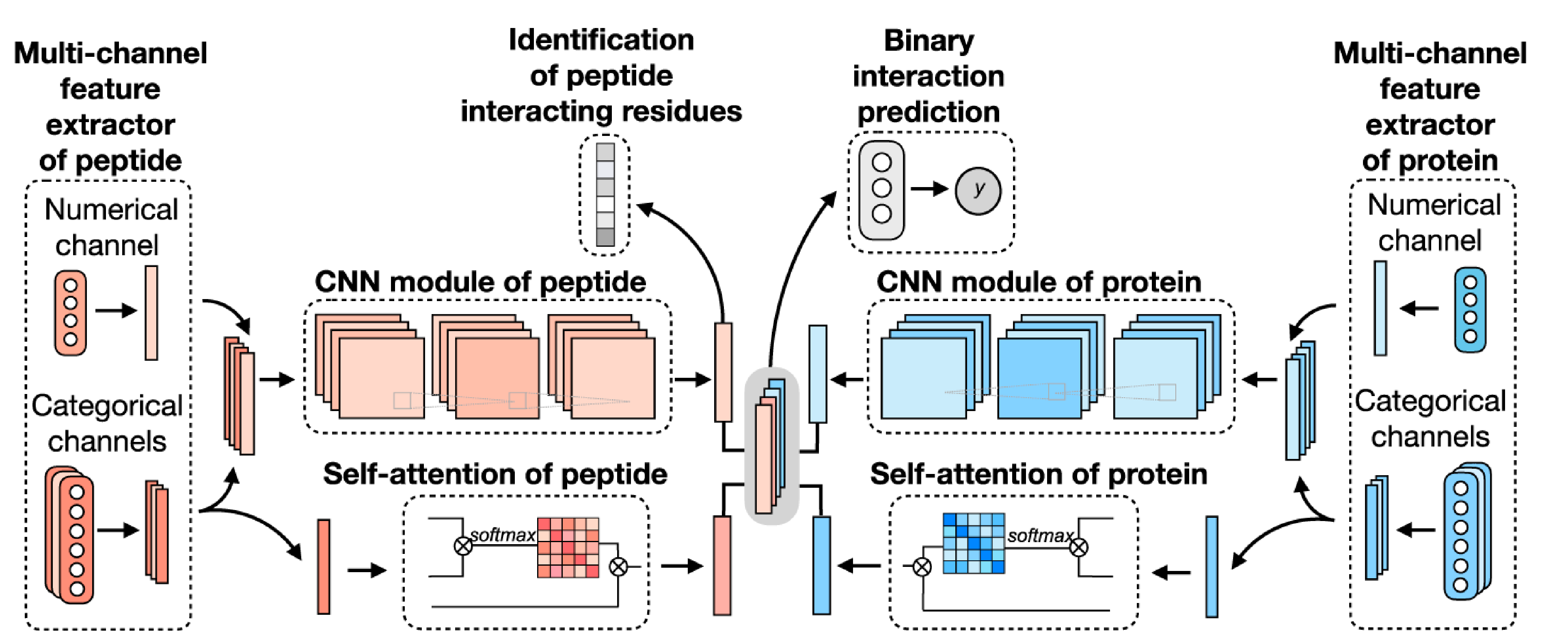}
   \caption{The network architecture of CAMP. For each protein-peptide pair, the numerical and categorical features of peptide and protein sequences are extracted and fed into CNN modules. The outputs of the amino acid representations of the peptide and protein are also fed into the self-attention modules to learn the importance of individual residue to the final prediction. Then taking the outputs of CNN and self-attention modules together as input of three fully connected layers to predict the a binding score for each peptide-protein pair. The output of CNN modules is also used for predicting a binding score for each residue from peptide sequence. Adapted with permission from Y.Lei, S.Li, Z.Liu, F.Wan, T.Tian, S.Li, D.Zhao and J.Zeng, \textit{Nature Communications}, 2021, \textbf{12}, 5465. This article is licensed under the Creative Commons CC BY license, permitting unrestricted reproduction and adaptation provided proper crediting to author and source. Copyright 2024 Springer Nature.}
   \label{fig:7}       
\end{figure}

CAMP extracted data from difference sources, including RCSB PDB \cite{Berman2000, Burley2018} and the known peptide drug-target pairs from DrugBank \cite{Wishart2006, Wishart2007, Knox2010, Law2013, Wishart2017}.  For each PDB complex, protein-ligand interaction predictor (PLIP) is employed to identify non-covalent interactions between the peptide and the protein, considering these interactions as positive samples for training. Additionally, PepBDB \cite{Wen2018} aids in determining the binding residues of peptides involved in the specific protein-peptide complexes. Various features are extracted based on their primary sequences to construct comprehensive sequence profiles for peptides and proteins. These features include secondary structure, physicochemical properties, intrinsic disorder tendencies, and evolutionary information \cite{Zhao2018, Mszros2018, Magnan2014, Madeira2019, Hamp2015}. CAMP utilizes two multi-channel feature extractors to process peptide and protein features separately (Figure \ref{fig:7}). Each extractor contains a numerical channel for numerical features (PSSM and the intrinsic disorder tendency of each residue), along with multiple categorical channels for diverse categorical features (raw amino acid, secondary structure, polarity and hydropathy properties). Two CNN modules extract hidden contextual features from peptides and proteins. Self-attention layers are also employed to capture long-range dependencies between residues and assess the contribution of each residue to the final interaction. CAMP applies fully connected layers on all integrated features to predict the interaction between proteins and peptides. In addition to binary interaction prediction, CAMP can identify which residue of peptides interacts with target proteins by adding a sigmoid activation function to the output of the peptide CNN module. Compared with three baseline models (DeepDTA \cite{ztrk2018}, PIPR \cite{Chen2019}, NRLMF \cite{Liu2016}), CAMP demonstrates consistent better performance with an increase by up to 10\% and 15\% in terms of Area Under the Curve (AUC) and Area Under the Precision-Recall Curve (AUPR). To evaluate its ability to identify binding residues of peptides, the predicted label of each residue of the peptide is compared with real label for four existing peptide binders. The results shows that CAMP correctly predicts binding residues and thus provides reliable evidence for peptide drug design. 

\begin{figure}[!htbp]
   \centering
   \includegraphics[width=\textwidth]{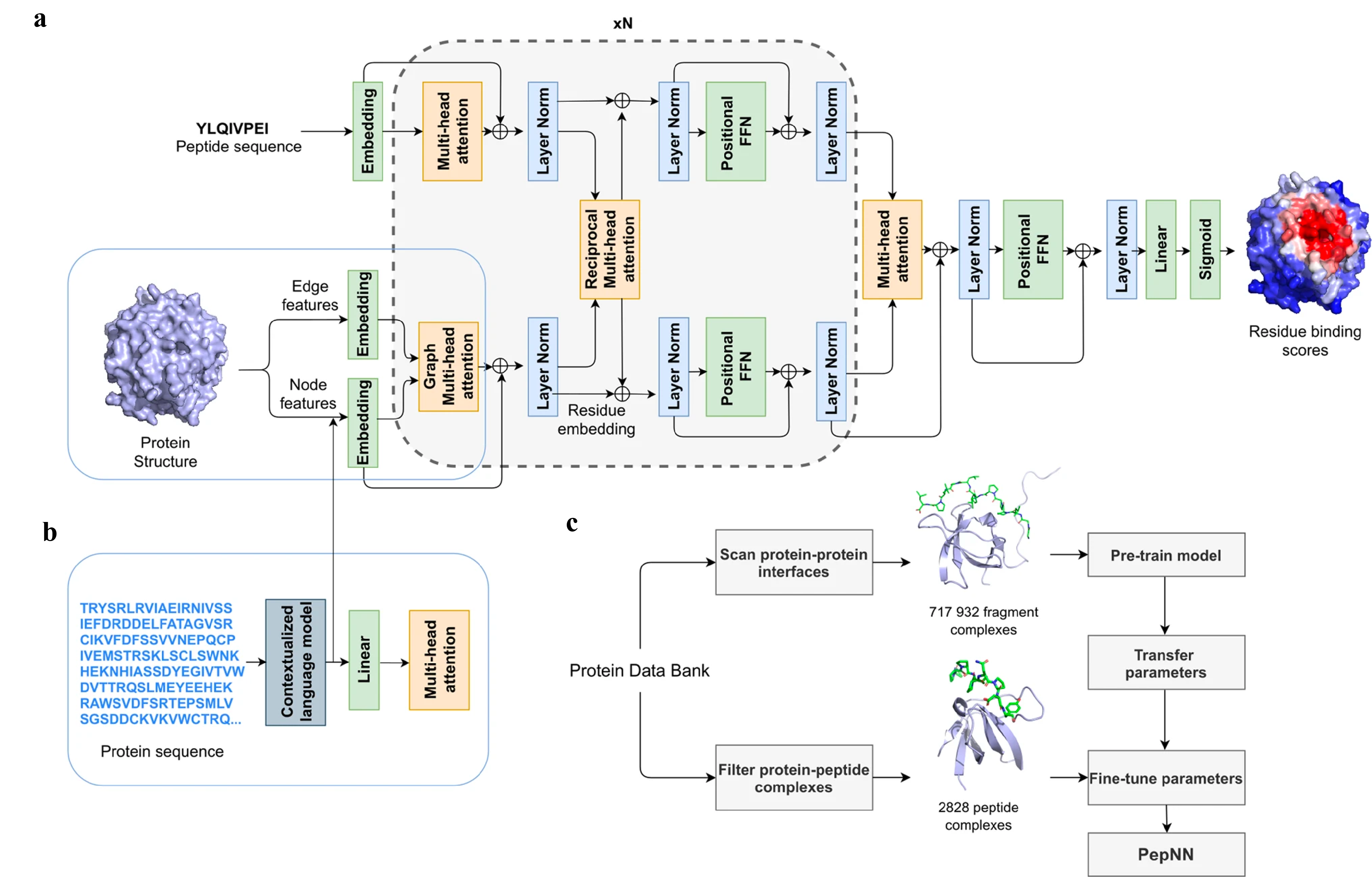}
   \caption{The model architecture and training procedure of PepNN. (a) The input of PepNN-Struct and model architecture. Attention layers are indicated with orange; normalization layers are indicated with blue and simple transformation layers are indicated with green. (b) The input of PepNN-Seq. (c) Transfer learning pipeline used for training PepNN. Reproduced with permission from O. Abdin, S. Nim, H. Wen and P. M. Kim, \textit{Communications Biology}, 2022, \textbf{5}, 503. This article is licensed under the Creative Commons CC BY license, permitting unrestricted reproduction and adaptation provided proper crediting to author and source. Copyright 2024 Springer Nature.}
   \label{fig:8}       
\end{figure}

Instead of only applying self-attention layer, Adbin \textit{et al.} developed a Transformer-based architecture known as PepNN, enabling both sequence-based (PepNN-Seq) and structure-based (PepNN-Struct) predictions of peptide binding sites \cite{Abdin2022}. PepNN takes representations of a protein and a peptide sequence as inputs and generates a confidence score for each residue,  indicating the likelihood of being part of binding sites. PepNN-Struct learns a contextual representation of a protein structure through the use of graph attention layers (Figure \ref{fig:8}a). In contrast, PepNN-Seq only takes the protein and peptide sequence as inputs (Figure \ref{fig:8}b). In the PepNN algorithm, the encoding of the peptide sequence is independent from the protein encoding module, under the assumption that the peptide sequence carries all the necessary information regarding peptide-protein binding. However, in many scenarios, the peptide sequence is not sufficient to determine the bound conformation, as the same peptide can adopt different conformations when bound to different proteins \cite{Mohan2006}. Motivated by this, PepNN incorporates a multi-head reciprocal attention layer that simultaneously updates the embeddings of both the peptide and protein (Figure \ref{fig:8}a). This module attempts to learn the interactions between protein and peptide residues involved in binding.

Another challenge in predicting the protein-peptide binding sites is the limited availability of protein-peptide complex training data. Protein-protein complex information was added to the training set to overcome the limited data issue. Notably, not entire protein-protein complex data was included, because the interactions between two proteins can be mediated by a linear segment in one protein that contribute to the majority of the interface energy. Pre-training of the model was conducted using a substantial dataset of large protein fragment-protein complexes (717,932) \cite{Sedan2016}. Fine-tuning of the model then took place with a smaller set of peptide-protein complexes (2,828), resulting in a considerable enhancement in predictive performance, particularly for the PepNN-Struct model (Figure \ref{fig:8}c). PepNN reliably predicts peptide binding sites on an independent test set and three benchmark datasets from the other studies \cite{ Zhao2018, Taherzadeh2017, Johanssonkhe2019}. PepNN-Struct surpassed most peptide binding site prediction approaches, achieving a higher AUC score. While PepNN generally exhibits lower MCC than the SOTA method AlphaFold-Multimer in most cases, its independence from multiple sequence alignments may render PepNN more suitable for modeling synthetic PepPIs.

While numerous computational methods have been developed for predicting peptide-protein binding site, many of them need complex data preprocessing to extract features, often resulting in reduced computational efficiency and predictive performance. Wang \textit{et al.} developed an end-to-end predictive model that is independent of feature engineering named PepBCL \cite{Wang2022PEPBCL}. This innovative approach leverages pre-trained protein language models to distill knowledge from protein sequences that are relevant to protein structures and functions. Another challenge encountered in identifying protein-peptide binding sites is the issue of imbalanced data. Current work typically construct a balanced dataset by using under-sampling techniques. However, these techniques remove samples from the majority class to match the size of minority class. In PepBCL algorithm, a contrastive learning-based module is introduced to tackle this problem. Unlike conventional under-sampling methods, the contrastive learning module adaptively learn more discriminative representations of the peptide binding residues. 

\begin{figure}[!htbp]
   \centering
   \includegraphics[width=\textwidth]{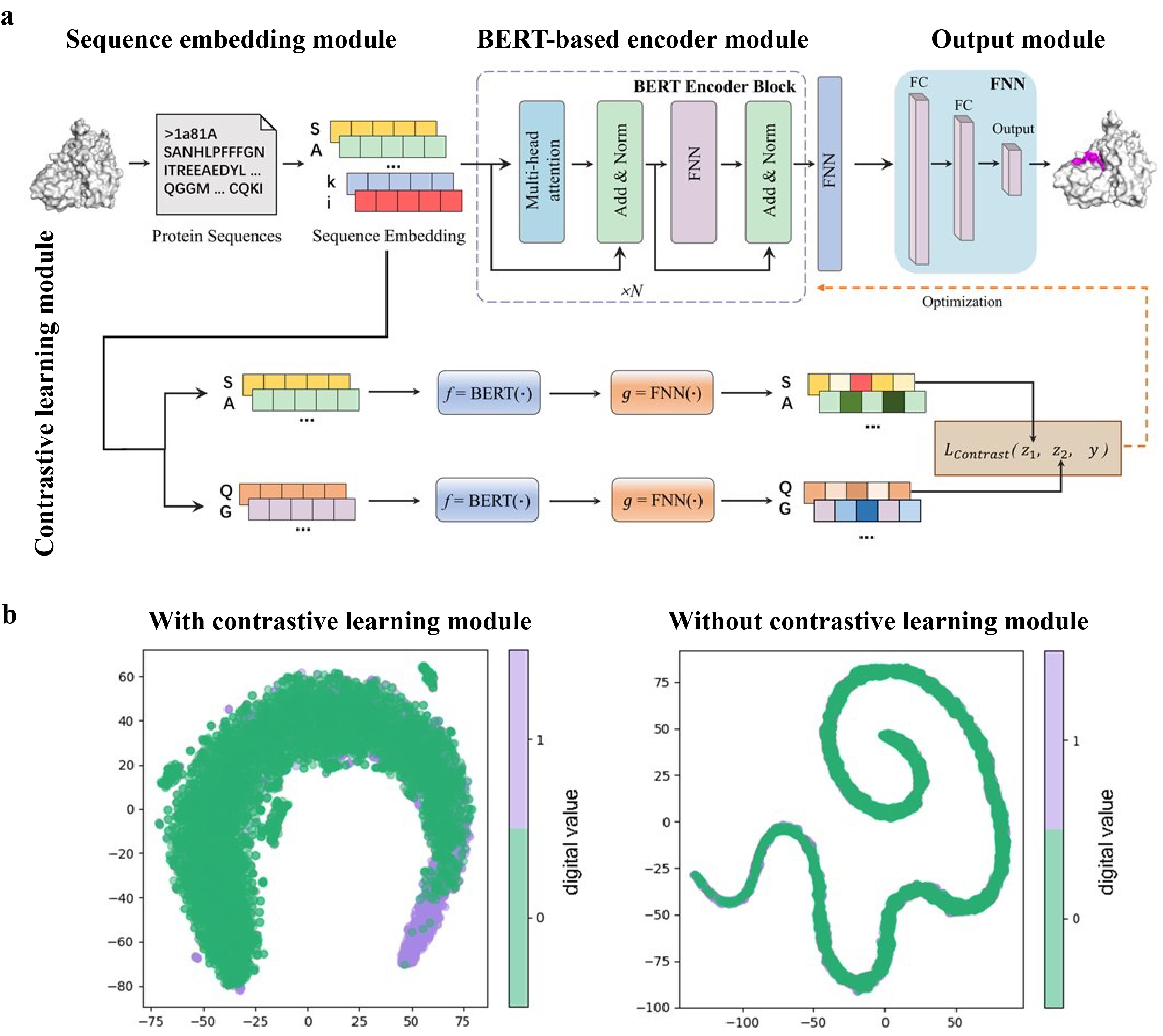}
   \caption{(a) Architecture of PepBCL consists of four modules. Sequence embedding module: convert protein sequence to sequence embedding for each residue; BERT-based encoder module: extract high-quality representations of each residue in protein; Output module: predict the label (binding/non-binding) of residues using fully connected layers; and contrastive learning module:  obtain more distinguishable representations by minimizing contrastive loss. (b) t-SNE visualization of the feature space distribution of PepBCL with/without contrast module on testing dataset. Reproduced with permission from R. Wang, J. Jin, Q. Zou, K. Nakai and L. Wei, \textit{Bioinformatics}, 2022, \textbf{38}, 3351–3360. Copyright 2024 Oxford University Press.}
   \label{fig:9}       
\end{figure}

The PepBCL architecture is composed of four essential modules: sequence embedding module, BERT-based encoder module \cite{https://doi.org/10.48550/arxiv.1810.04805}, output module and contrastive learning module \cite{pmlr-v119-chen20j, https://doi.org/10.48550/arxiv.1911.05722}. In the sequence embedding module, each amino acid of the query sequence is encoded into a pre-trained embedding vector, while the protein sequence is encoded to an embedding matrix. In the BERT-based encoder module, the output from the sequence embedding module undergoes further encoding through BERT to generate a high dimensional representation vector \cite{Elnaggar2022}. The representation vector is then passed through a fully connected layer. In the contrastive learning module, the contrastive loss between any two training samples is optimized to generate more discriminative representations of the binding residues. In the output module, the probability of each residue being in a binding site is calculated (Figure \ref{fig:9}a). When compared with the existing sequence-based method (SPRINT-Seq \cite{Taherzadeh2016}, PepBind \cite{Zhao2018}, Visual \cite{Wardah2020}, and PepNN-Seq \cite{Abdin2022}), PepBCL achieves a significant improvement in the precision by 7.1\%, AUC by 2.2\%, and MCC by 1.3\% over best sequence predictor PepBind \cite{Zhao2018}. Furthermore, PepBCL also outperforms all structure-based methods (i.e. Pepsite \cite{Petsalaki2009}, Peptimap \cite{Lavi2013}, SPRINT-Str \cite{Taherzadeh2017}, and PepNN-Struct \cite{Abdin2022}) in terms of MCC. The superior performance of PepBCL indicates that DL approaches can automatically learn features from protein sequence to distinguish peptide binding residues and non-binding residues, eliminating the reliance on additional computational tools for feature extraction. When assessing various methods using evaluation metrics, it is observed that recall and MCC tend to be notably low due to the extreme class imbalance in the dataset. This suggests that many true protein-peptide binding residues may be overlooked. However, PepBCL demonstrates improved recall and MCC values, highlighting the effectiveness of the contrastive module in identifying more true peptide binding residues. This enhancement can be attributed to the contrastive learning's ability to extract more discriminative representations, particularly in imbalanced datasets. Figure \ref{fig:9}b visually demonstrates the learned feature space with and without the contrastive learning module, showcasing a clearer distribution of binding and non-binding residues in the feature space.

\textbf{AlphaFold/RoseTTAFold/OmegaFold/ESMFold}. Multiple Sequence Alignment (MSA)-based transformer models such as AlphaFold2 (AF2, including monomer model \cite{Jumper2021} and multimer model \cite{Evans2021}), RoseTTAFold,\cite{Baek2021} and protein Language Model (pLM)-based models such as OmegaFold,\cite{Wu2022} and ESMFold,\cite{Lin2023} have demonstrated remarkable success in predicting the \textit{in silico} folding of monomeric proteins and peptides.\cite{McDonald2023} However, PepPIs are relatively flexible protein complexes, making it challenging to achieve highly accurate predictions. Therefore, benchmarking these SOTA DL techniques on PepPI predictions could provide structural insights into peptide–protein complexes, for example, binding affinities, conformational dynamics, and interaction interfaces, thus contributing to the advancement of molecular biology and drug discovery.

\begin{figure}[!htbp]
   \centering
   \includegraphics[width=0.8\textwidth]{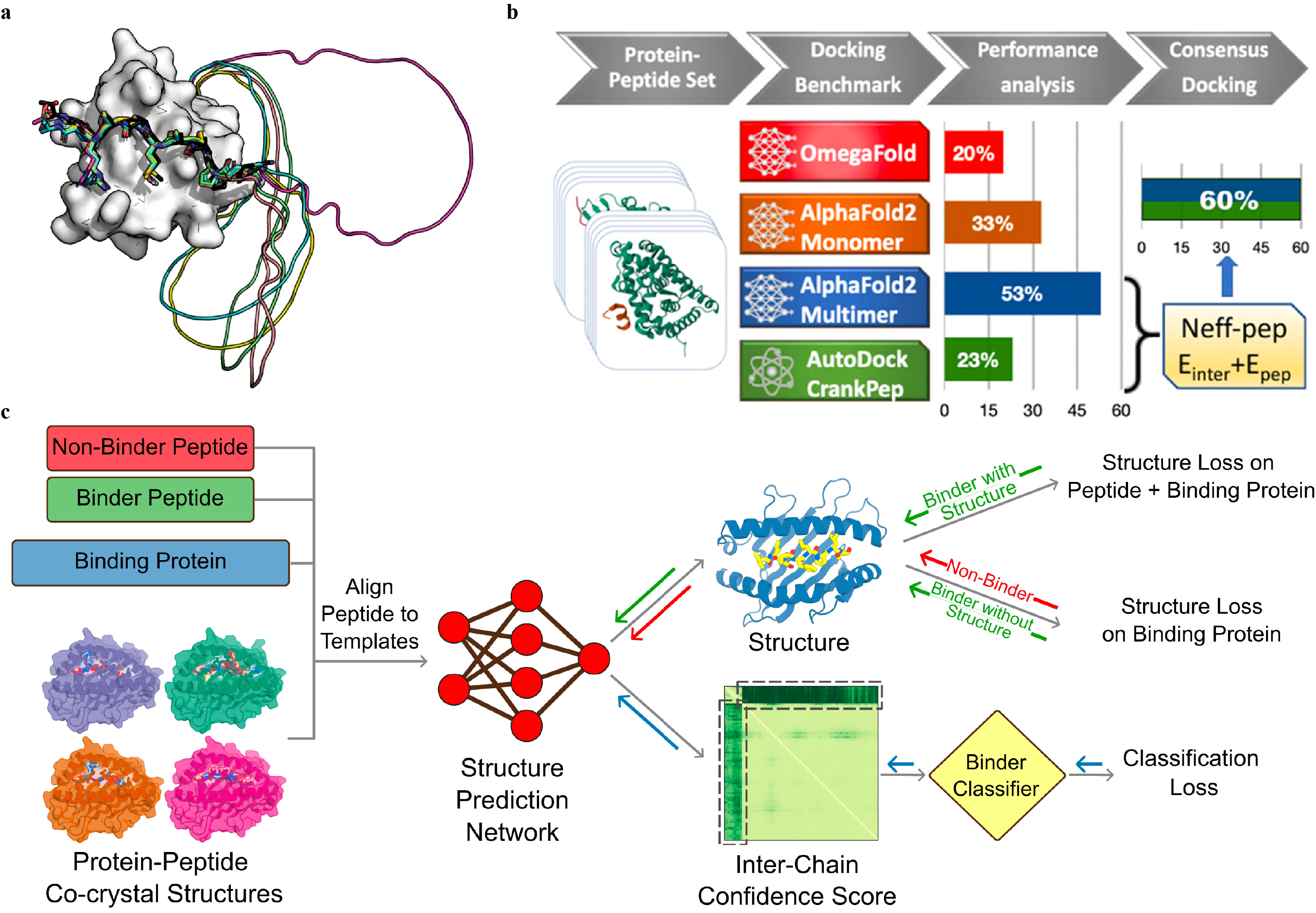}
   \caption{(a) A successful example (PDBID: 1SSH) of peptide–protein docking with a poly-glycine linker via AlphaFold2. This method can dock the peptide at the correct position (native peptide is shown in black, docking ppeptides are shown in other colors) and identify the linker as unstructured region (modeled as a circle). Adapted with permissions from T. Tsaban, J. K. Varga, O. Avraham, Z. Ben-Aharon, A. Khra-mushin and O. Schueler-Furman, \textit{Nature Communications}, 2022, \textbf{13}, 176, this article is licensed under the Creative Commons CC BY license, permitting unrestricted reproduction and adaptation provided proper crediting to author and source. Copyright 2024 Springer Nature. (b) AlphaFold2-Multimer model outperforms other DL approaches and achieves remarkable docking success rates of 53\% for peptides-protein docking. A designed docking approach combining ADCP and AlphaFold2-Multimer achieves an improved success rates of 60\%. Adapted with permissions from S.Shanker and M.F.Sanner, \textit{Journal of Chemical Information and Modeling}, 2023, \textbf{63}, 3158–3170. Copyright 2024 American Chemical Society.  (c) Mechanism of structure prediction networks for peptide binder classification by fine-tuning AlphaFold2. The input of the model includes the peptide binder and non-binder sequences, protein sequences, and peptide-protein co-crystal structures as templates. After positionally aligning the peptide sequence to the template, the complex structure is then predicted with AlphaFold2. A binder classification layer converts the AlphaFold2 output PAE values into a binder/non-binder score. The combined loss function including the structure loss over the entire complex for peptide binder and over protein only for non-binder, and classification loss from the binder classification layer, is used for model training.  Adapted with permissions from A. Motmaen, J. Dauparas, M. Baek, M. H. Abedi, D. Baker and P. Bradley, \textit{Proceedings of the National Academy of Sciences}, 2023, \textbf{120}, e2216697120. This article is licensed under a Creative Commons Attribution 4.0 (CC BY) License, permitting unrestricted reproduction and adaptation provided proper crediting to author and source. Copyright 2024 National Academy of Science.}
   \label{fig:10}       
\end{figure}

While AF2 monomer was originally designed for predicting monomeric proteins/peptides structures, it has recently been shown to be successful in predicting PepPIs by Tsaban et al.\cite{Tsaban2022} The PepPIs could be represented as the folding of a monomeric protein by connecting the peptide to the C-terminus of the receptor with a poly-glycine linker (Figure \ref{fig:10}a), which forms a general idea of how to perform peptide–protein docking using the AF2 monomer model. This method can not only identify the peptide binding regions but also accommodate binding-induced conformational changes of the receptor. AF2 surpassed RoseTTAFold since the latter tended to fold the polyglycine linker into a globular structure or various interactive loops. For a small dataset of 26 PepPI complexes, AF2 achieved a relatively high accuracy (75\%) for complexes whose binding motifs have been experimentally characterized. AF2 also outperformed another peptide docking method PIPER-FlexPepDock (PFPD) \cite{Alam2017} in terms of both accuracy and speed. Furthermore, accurate predictions were achieved with AF2 pLDDT values above 0.7, further verifying that AF2 monomer can reliably predict the PepPIs. However, the predicted accuracy became lower (37\%) when tested on a larger dataset (96 complexes), indicating that further improvements are needed for more accurate PepPI predictions by AF2 monomer. 

The recent release of AF2 multimer has yielded a major improvement in PepPIs prediction. Using a set of 99 protein–peptide complexes, Shanker \textit{et al} \cite{Shanker2023} compared the performance of AF2 monomer, AF2 multimer, and OmegaFold on PepPI prediction with their peptide docking software AutoDock CrankPep (ADCP).\cite{Zhang2019} The new AF2 multimer model with 53\% accuracy, which was trained to predict the interfaces of multimeric protein complexes, outperformed OmegaFold with 20\% accuracy and ADCP with 23\% accuracy (Figure \ref{fig:10}b). However, the AF2 multimer model is only limited to linear peptides, reducing its applicability to cyclized peptides, or peptides with non-standard amino acids. Effective selection from top-ranked poses yielded by both AF2 multimer and ADCP docking tool was found to further enhance the accuracy to 60\%. Therefore, DL protein structure prediction models, especially AF2 multimer, have achieved high-accuracy in PepPIs predictions, though limitations exist. Combining these SOTA DL models with traditional peptide docking tools could be a future direction for further improving the accuracy of PepPIs predictions.

Leveraging the highly accurate predictions of protein structures by AF2, Amir Motmaen \textit{et al} \cite{Motmaen2023} developed a more generalized model for the prediction of PepPIs. The model was accomplished by placing a classifier on top of the AF2 network and fine-tuning the combined network (Figure \ref{fig:10}c). AF2 was able to achieve optimal performance and generate the most accurate complex predicted structure models for a large dataset of peptide-Major Histocompatibility Complex (MHC) complexes. This was accomplished by aligning the peptide sequence with the peptide-protein crystal structures as templates. However, AF2 occasional docking of non-binding peptides in the peptide binding domain of MHC highlighted the need for a clear classification of binder and non-binder peptides in the training of the model. To address this issue, a logistic regression layer that normalizes AF2 Predicted Aligned Error (PAE) score into binder/non-binder score was placed on top of AF2. This resulted in three types of losses being combined and applied to further fine-tune the combined model: structure loss on both peptide and protein for binding peptide-protein complexes, structure loss on protein only for non-binding peptide-protein complexes, and classification loss on binding/non-binding score. The evaluation of the combined model showed a ROC AUC of 0.97 for Class \MakeUppercase{\romannumeral 1}  and 0.93 for Class \MakeUppercase{\romannumeral 2} peptide-MHC interactions. Surprisingly, the fine-tuned model outperformed the previously mentioned HSM model and could also be generalized on PDZ domains (C-terminal peptide recognition domain) and SH3 domains (proline-rich peptide binding domain), despite being trained and fine-tuned only on the peptide-MHC dataset. Therefore, taking advantage of the accurate predictions of protein structures through AF2, and fine-tuning the model with existing peptide-protein binding data offers significant boost to PepPIs predictions. 

\section{Conclusions and Future Research Directions}

Peptides, which are short proteins consisting of around 2 to 50 amino acids, are known for their flexibility. This characteristic makes it challenging to achieve highly accurate predictions of PepPIs. A variety of SOTA ML and DL models summarized in this review have been designed and applied to predict PepPIs, which are key to \textit{de novo} peptide drug design. 

Apart from their well-documented high efficiency and accuracy requirements, ML/DL methods offer several other advantages in the predictions of PepPIs. Compared to Docking or MD Simulation methods, ML or DL methods offer diverse options for model inputs. DL methods, such as transformers and language models, have been shown to achieve great success in predicting PepPIs solely on sequence information. Instead of original sequence or structure information, ML methods can also incorporate multi-level information such as evolutionary information, secondary structures, solvent accessible surface area, and so forth, which could significantly enhance the accuracy of the prediction. Furthermore, more interpretability can be provided by ML/DL methods. Attention mechanism assists in demonstrating the internal dependencies between residues and the contribution of each residue to PepPIs. Graph models capturing multi-scale structure information of peptides and proteins are able to provide insights into the underlying peptide-protein binding chemical principles and binding patterns. Moreover, ML/DL techniques exhibit a degree of generalizability. Some advanced techniques like transfer learning or one-shot learning models, which have been applied in protein engineering and protein-ligand interaction prediction \cite{Shamsi2020,Horne2022, AltaeTran2017, Mi2022}, could facilitate the models trained on certain peptide-protein binding datasets to generalize to other peptide-protein complexes.
 
Despite their numerous advantages, ML and DL methods also have certain limitations in the prediction of PepPIs, which highlight potential areas for future research. One significant challenge is the issue of imbalanced datasets in the training and testing of PepPIs prediction models. Given that peptide binding is typically a rare occurrence, the imbalanced number of positive and negative samples often results in the limited performance of ML/DL models due to the poor understanding of the minority binding class. Consequently, ML/DL methods for PepPI predictions were normally trained based on datasets with positive-to-negative ratio as 1:1. Both oversampling methods, which duplicate or create new samples, and undersampling methods, which delete or merge samples in the majority class can enhance the model performance on imbalanced classification. Besides, challenges arise when dealing with peptides deeply embedded in the enzyme's active site especially involving cofactors. Accurate predictions for such interactions require high-quality structural training data reflecting correct folding for both peptide and enzyme along with the precise knowledge of buried peptide binding positions and poses. Furthermore, accurate geometric and electronic considerations of cofactors would be necessary to predict the peptide and protein residue interactions with the co-factors. The scarcity of structural training data for such instances results in a relatively worse model performance on PepPIs. Recent efforts, such as RoseTTAFold All-Atom\cite{Krishna2023} (RFAA), aim to address this challenge. RFAA can model full biological assemblies, including metal cofactors, by training on a comprehensive dataset comprising sequence information, residue pairwise distance from homologous templates, and coordinates of protein-small molecule, protein-metal, and covalently modified protein complexes. As a result, RFAA demonstrates reasonable prediction performance and stands out as the first model capable of predicting arbitrary higher-order biomolecular complexes, encompassing multiple proteins, small molecules, metal ions, and nucleic acids. However, this is a recent development so there are no applications of RFAA to PepPIs prediction. As advancements in structural biology and computational methods continue, it is foreseeable that more sophisticated models will emerge, further enhancing the capability to accurately predict PepPIs, even involving buried peptides and cofactors. Additionally, ML/DL methods often failed in the prediction of PepPIs between intrinsically disordered peptides (IDP) and proteins. IDPs are abundant in nature, with flexible and disordered structures but adopt stable and well-defined structures upon binding. In these cases, ML/DL methods, particularly structure-based models, tend to fail in predicting binding sites and peptide binding conformations, offering little insights into the binding mechanism. With the enhancement of computing power, high-throughput MD simulations can achieve more accurate predictions of binding sites and peptide/protein conformations as well as a deeper understanding of the mechanism of folding and binding, induced fit (binding then folding), or conformational selection (folding then binding).  The integration of MD or quantum chemical insights and ML/DL methods could constitute a promising future research direction of PepPIs predictions. 

Another future direction is to develop ML/DL models to predict cyclic peptide and protein interaction. Cyclic peptides have emerged as a promising therapeutical modality because of distinct pharmacological characteristics in comparison to small molecules and biologics \cite{Tsomaia2015, Vinogradov2019, Muttenthaler2021}. For example, cyclic peptides are more resistant to digestive enzymes like peptidases and exoproteases due to their stable cyclic structures. Cyclic peptides have a broader interaction surface than small-molecule drugs and thus may function as inhibitors with high affinity and selectivity for modulating protein-protein interactions. Furthermore, cyclic peptides exhibit better permeability across cell membranes and less expensive to synthesize compared to antibodies. However, the development of deep learning models for designing cyclic peptides has faced challenges, mostly due to the small number of available structures. Recently, Rettie \textit{et al.} introduced the AfCycDesign approach, a novel modification of AlphaFold network for accurate structure prediction and design of cyclic peptides \cite{Rettie2023}. Standard positional encoding in AlphaFold is based on the position of each amino acid in the linear peptide, with the termini being the maximum distance from each other. AfCysDesign modifies the positional encoding with cyclic offset such that the termini are connected to each other. This approach can accurately predict the structures of cyclic peptides from a single sequence, with 36 out of 49 cases predicted with high confidence (pLDDT $>$ 0.85) matching the native structures with root mean squared deviation (RMSD) $<$ 1.5 $\AA$. Kosugi \textit{et al.} employed the relative positional encoding with cyclic offset to predict protein-cyclic peptide complexes \cite{Kosugi2023}. The cyclic offset was only applied in the cyclic peptide region, while the positional encoding of protein region remained the default one. The predictions outperformed state-of-the-art local docking tools for cyclic peptide complexes.  

Future research directions should also prioritize the enhancement of model's ability to generate novel peptide sequences to specific target proteins of interest, thereby contributing to \textit{de novo} peptide drug design. An essential way is to fine-tune pre-trained pLM. Introducing noises and perturbations within the peptide latent space of pLM, or masking peptide sequences to facilitate the model to learn the probability distribution of peptide binders, could be explored to generate entirely new peptide sequences. Additionally, diffusion models offer another avenue for achieving the generative tasks. These models possess a deeper understanding of the intricate molecular interactions at the atomic levels, thus enabling the generation of new peptide sequences based on peptide-protein complex structures. The resultant novel peptide sequences can be subsequently validated through MD simulations, \textit{in vitro}, and \textit{in vivo} experimental tests. Therefore, developing new generative models or leverage the pre-trained ML/DL models to facilitate peptide generation represents a noteworthy and promising future for advancing peptide drug design.

In conclusion, ML/DL-guided methods have shown significant potential for the accurate predictions of peptide-protein complex structures and binding sites. These SOTA models will undoubtedly further accelerate the process of peptide drug discovery and design.

\begin{acknowledgement}
D.S. acknowledges support from National Institutes of Health, under Award No. R35GM142745 and No. R21AI-167693.
\end{acknowledgement}


\bibliography{PepPI_review_ref}

\end{document}